\documentclass[twocolumn,amssymb, showpacs, amsmath,prb]{revtex4}
\usepackage{graphicx}
\usepackage{dcolumn}
\usepackage{bm}

\begin{document}

\title{Simulation of the current dynamics in superconductors: Application to magnetometry
measurements}

\author {M. Zehetmayer}
\affiliation {Vienna University of Technology, Atominstitut, 1020
Vienna, Austria}
 \email{zehetm@ati.ac.at}

\date{\today}

\begin{abstract}
A simple model for simulating the current dynamics and the magnetic properties of superconductors is
presented. Short simulation times are achieved by solving the differential form of Maxwell's
equations inside the sample, whereas integration is only required at the surface to meet the exact
boundary conditions. The procedure reveals the time and position dependence of the current density
and the magnetic induction ($B$) making it very convenient to apply field dependent material
parameters for the simulation of magnetization loops, relaxation measurements, etc. Two examples,
which are important for standard magnetometry experiments, are discussed. Firstly, we prove that
evaluating the critical current density ($J_{\text{c}}$) from experiment by applying Bean's model
reveals almost the exact $J_{\text{c}}(B)$ behavior, if the evaluation is corrected by a simple
numerical expression. Secondly, we show that the superconducting volume fraction of a sample can be
directly determined from magnetization loops by carefully comparing experiment and simulation in the
field range, where the current loops are differently oriented within the sample.
\end{abstract}

\pacs{74.25.Ha,74.25.Sv,02.60.Cb}

\maketitle

\section{Introduction}

Simulating the magnetic behavior of type II superconductors requires solving Maxwell's equations
under specific material equations. Many different models have been proposed for solving this task
including energy minimization, variational, or more direct approaches (e.g. Refs.
\onlinecite{Pri96a,Bra96a,Bra99a, Lab97a,San01b, Bad02a,Wol08,Cam09a,Lu08a}). When addressing not
only the static, but also the dynamic behavior, it is often convenient to calculate the time
evolution of field relevant quantities (e.g. of the magnetic vector potential or of the induction)
directly.

Still, different approaches can be chosen. The most direct way is to solve Maxwell's equations by
integration (i.e. with the help of the appropriate Green's function) which yields, e.g., the
current density or the vector potential.\cite{Bra99a} This method naturally leads to results
fulfilling the electromagnetic boundary conditions at the sample surface. However, integrating over
the whole sample volume requires a lot of calculation time and is usually too slow for
three-dimensional (3D) problems. For a grid of $N^3$ elements, into which the sample is
divided for the numerics, at least $N^6$ operations have to be performed per time step.

The opposite approach deals with the differential form of Maxwell's equations, where fields and
currents are locally related by difference equations.\cite{Lu08a} This makes the calculations very
fast inside the sample, since only about $N^3$ operations per time step are required. Care has to be
taken, however, with the boundary conditions at the surface, which requires solving the
equations also outside the sample up to distances, where the influence of the sample can be
neglected.

As a compromise, we present a model, where the differential equations are applied to the sample
interior, whereas the values at the sample boundaries are obtained by integrating over the
sample volume yielding the correct boundary conditions. This requires $N^3 + 6 N^4$
operations. It was already demonstrated\cite{Zeh06a} that the model worked sufficiently rapidly for
3D calculations on a standard PC.

In this paper, we provide details of the method and discuss examples relevant for every-day
magnetometry measurements. In section \ref{sec:simulation} the simulation procedure and some
technical aspects of the implementation are described. We start with a general model, but
concentrate in the following on rectangular samples, for which several tests were carried out.
Section \ref{sec:application} is devoted to applications of the method to magnetometry (e.g. SQUID
or VSM) measurements. We will verify the well known experimental evaluation of the critical current
density from magnetization loops by applying Bean's model and introduce a simple way of determining
the superconducting volume fraction of a sample.

\section{Simulation Model \label{sec:simulation}}

Our aim is to solve Maxwell's equations relevant for the magnetic properties of a superconductor. We
assume superconductors without (non-superconducting) para- or diamagnetic properties and initially
ignore the reversible magnetization of the superconducting state. Thus, Maxwell's equations (in SI
units) for vacuum may be applied.
\begin{eqnarray}
 	\label{eqn:Maxwell1}
	\vec{\nabla} \cdot \vec{B} &=& 0 \\
	\label{eqn:Maxwell2}
	\vec{\nabla} \times \vec{B} &=& \mu_0 \vec{J} \\
	\label{eqn:Maxwell3}
	\partial_{\rm t} \vec{B} &=& - \vec{\nabla} \times \vec{E}
\end{eqnarray}
The displacement current (Eq.~\ref{eqn:Maxwell2}) is not considered. $\vec{B}$ denotes the magnetic
induction, $\mu_0 = 4 \pi \times 10^{-7}$\,TmA$^{-1}$ the vacuum permeability, $\vec{J}$ the
current density and $\vec{E}$ the electric field. Generally, the vector quantities depend on time
($t$) and position ($\vec{r}$). The electric field is usually explicitly provided, e.g. as a
function of current density, induction, and material specific parameters such as the critical
current density to specify the superconducting properties of the material. It is directed parallel
to $\vec{B} \times \vec{v}$, where $\vec{v}$ denotes the velocity of the vortices caused by the
Lorentz force $\vec{J} \times \vec{B}$. In the following, we will address rather simple
configurations, relevant for most magnetometry experiments, where the currents flow perpendicular to
$\vec{B}$ and thus $\vec{E}$ and $\vec{J}$ are parallel. The more complicated situation, where
$\vec{J}$ has also a (usually small) component parallel to $\vec{B}$, which can occur in the case of
arbitrary field directions or irregular sample geometries, etc.,  is not considered. Recently, this
problem was solved for simple configurations in the (static) Bean limit \cite{Mik05a,Bra07a}, but it
is currently not establish how to obtain more general results. For the simpler examples, discussed
here, different kinds of material equations are available in the literature, which describe
different materials and experimental situations (e.g. Refs.~\onlinecite{Bla94a,Bra95a,Yes96a}). Most
prominent is the power law
\begin{equation}
 	\label{eqn:PowerLaw}
	E = E_{\rm c} \left(  \frac{J}{J_{\text{c}}} \right)^n
\end{equation}
which reproduces the flux creep behavior of type II superconductors quite well. Here, $E_{\rm c}$ is
the electric field criterion, $J_{\text{c}}$ the critical current density, and $n$ the so-called
$n$-value, which is a two-digit number in most cases. Very large $n$ values (e.g. $n = \infty$)
reproduce the Bean model\cite{Bea62a} ($E = 0$ for $J < J_{\text{c}}$ and $E = \infty$ for  $J >
J_{\text{c}}$). Both $J_{\text{c}}$ and $n$ may depend on $\vec{B}$ and other parameters.

\subsection{Procedure}

\begin{itemize}
 \item[i.] Our starting point is a convenient initial state of the magnetic induction
$\vec{B}(\vec{r}, t)$ and the corresponding current density $\vec{J}(\vec{r}, t)$ (satisfying
$\vec{\nabla} \times \vec{B} = \mu_0 \vec{J}$). Usually, we start with $\vec{B} = 0$ and $\vec{J} =
0$ in the whole sample or with the results of a previous simulation.

\item[ii.] Next, we calculate the electric field - $\vec{E}(\vec{r}, t)$ - by applying the material
equation (e.g. Eq.~\ref{eqn:PowerLaw}) and

\item[iii.] its curl, which corresponds to the time derivative of the induction ($\partial_{\rm t}
\vec{B}$, cf. Eq.~\ref{eqn:Maxwell3}) and thus leads to

\item[iv.]  the induction of the next time step
\begin{equation}
 	\label{eqn:Binterior}
	\vec{B}(\vec{r}, t + \Delta t) = \vec{B}(\vec{r}, t) + \partial_{\rm t} \vec{B}(\vec{r}, t)
\Delta t
\end{equation}
where $\Delta t$ denotes a small time increment. Equations (\ref{eqn:Maxwell3}) and
(\ref{eqn:Binterior}) are only applied to the interior of the sample, whereas

\item[v.] $\vec{B}(\vec{r}, t)$ at the sample surface is obtained by integrating over the sample
volume to fulfill the boundary conditions
\begin{equation}
 	\label{eqn:Bsurface}
	\vec{B}(\vec{r},  t + \Delta t) = \mu_0 [\vec{H}_{\text{a}}(\vec{r},  t + \Delta t) +
\vec{H}_{\text{s}}(\vec{r}, t)]
\end{equation}
with
\begin{equation}
 	\label{eqn:Hsurface}
	\vec{H}_{\rm s}(\vec{r}, t) = \frac{1}{4 \pi} \int\!d^3 {\rm r'} \frac{\vec{J}(\vec{r}', t)
\times
(\vec{r} -
\vec{r}')}{|\vec{r} - \vec{r}'|^3}
\end{equation}
$\vec{H}_{\text{a}}$ is the applied field. Equation~(\ref{eqn:Bsurface}) is the only place where
this quantity appears (i.e. $\vec{H}_{\text{a}}$ influences our calculations only via the surface).

\item[vi.] Knowing $\vec{B}$ in the whole sample allows evaluating $\vec{J}(\vec{r},t + \Delta t)$
from Eq.~(\ref{eqn:Maxwell2}). Note that $\vec{B}(\vec{r},  t + \Delta t)$ (\ref{eqn:Bsurface}) is
calculated from $\vec{H}_{\text{s}}(\vec{r},  t)$, which actually refers to the current distribution
of the previous time step. This calculation could be repeated with the new currents
$\vec{J}(\vec{r},  t + \Delta t)$, but this was found to be unnecessary due to the small value of
$\Delta t$.

\item[vii.] We end up with a new set of $\vec{B}$ and $\vec{J}$, close our simulation loop and
proceed with calculating the next $\vec{E}$ (step ii).
\end{itemize}

\begin{figure}
    \centering
    \includegraphics[clip, width = 8.5cm]{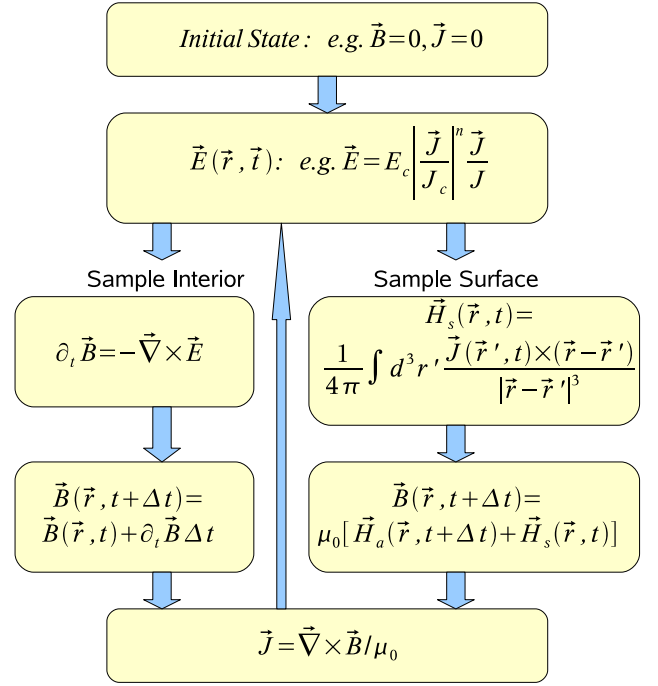}
    \caption{\label{fig:flow} (Color online) Flow diagram of the simulation sequence when only the
{\itshape irreversible} magnetization is taken into account (see the text for details).}
\end{figure}

Figure \ref{fig:flow} outlines the major parts of the sequence. After the last step (vii) specific
quantities may be evaluated (which is usually not done after each simulation loop, but after
predefined field or time intervals), e.g. the magnetic moment
\begin{equation}
 	\label{eqn:moment}
	\vec{m} = \frac{1}{2} \int\!d^3r\, \vec{r} \times \vec{J}
\end{equation}

Thanks to the use of the difference equations, the simulations can be carried out quite rapidly.
Time-consuming integration over the whole sample volume is only necessary for the induction at the
sample surface to provide the correct boundary values.

We wish to point out another very attractive feature of this method, namely the fact, that the
induction $\vec{B}$ is directly assessed in the simulations. This makes it very convenient to apply
any field dependent property - in particular a field dependent critical current density
$J_{\text{c}}(\vec{B})$ - without the need of any additional calculations of $\vec{B}$ (as would be
necessary for many other approaches).

Different specifications of $\vec{E}$ and $\vec{H}_{\text{a}}$ allow dealing with different
materials and experiments. For instance, the simple power law of Eq.~(\ref{eqn:PowerLaw}) leads to
the typical flux creep behavior of superconductors. A transport current can be considered by adding
external electric fields.

So far, only the irreversible properties have been addressed in the simulation procedure. A
simple way of taking also the reversible magnetization of a superconductor into account is presented
in Ref.~\onlinecite{Bra99a}, which can be easily adapted in our method by adding two further steps
to the above procedure.
\begin{itemize}
 \item[viii.] After finishing the last step (vii) of the original procedure, we know $\vec{B}$ and 
$\vec{J}$ in the entire sample. $\vec{B}$ is the overall local magnetic induction and thus includes
also the contribution from the reversible magnetization $\vec{M}_\text{r}$, i.e.
\begin{equation}
 	\label{eqn:B_Material}
	\vec{B} = \mu_0 (\vec{H} + \vec{M}_{\rm r}).
\end{equation}
The reversible behavior  may be assessed from Ginzburg Landau theory or from simpler
approximations (like those of Ref.~\onlinecite{Bra03a}), which provide $\vec{M}_{\rm r}$
as a function of $\vec{B}$ (or $\vec{H}$) and two parameters such as $B_\text{c2}$ - the upper
critical field - and $\kappa$ - the Ginzburg Landau parameter. Note that $\vec{M}_{\rm r}$ vanishes
at the sample surface.

 \item[ix.] Equation~(\ref{eqn:Maxwell2}) is still valid and the resulting $\vec{J}$ needed for 
calculating the boundary values (Eq.~(\ref{eqn:Hsurface})). Note that $\vec{J}$ includes also the
reversible part of the currents and thus may be particularly large at the sample surface. However,
 $\vec{J}$ does not enter the Lorentz force density ($f_\text{L}$) that drives the flux lines,
but\cite{Lab97a} $f_\text{L} = \vec{J}_\text{H} \times \vec{B}$, with
\begin{equation}
 	\label{eqn:jH}
	\vec{\nabla} \times \vec{H} =  \vec{J}_\text{H},
\end{equation}
where $\vec{H}$ is obtained from Eq.~(\ref{eqn:B_Material}). Accordingly, $\vec{J}_\text{H}$ (but
not $\vec{J}$) generates the electric field and we have to replace $\vec{J}$ by $\vec{J}_\text{H}$
in the expression for $\vec{E}$ (i.e. in the material equation). If the reversible magnetization is
ignored (as, e.g., in the flow chart of Fig.~ \ref{fig:flow}), $\vec{M}_{\rm r} = 0$ and thus
$\vec{J} = \vec{J}_\text{H}$.
\end{itemize}

\begin{figure}
    \centering
	 \includegraphics[clip, width = 8.5cm]{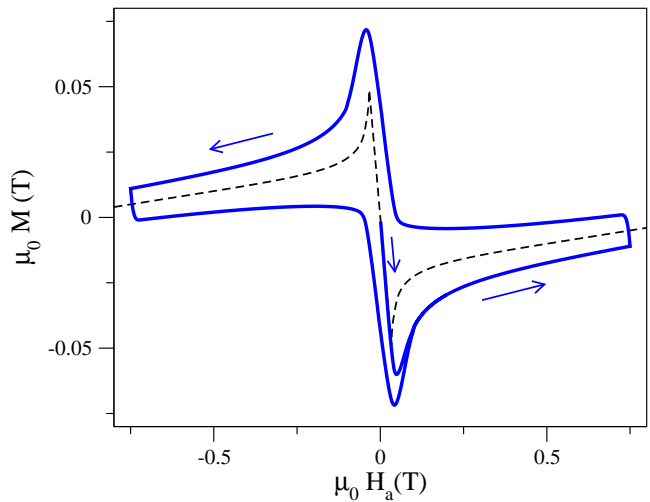}
    \caption{\label{fig:mrevirr} (Color online) Simulation of the magnetization (solid line, $M =
m/$volume) of a superconductor ($a \times b \times c = 1 \times 1 \times 1$\,mm$^3$), where both the
reversible and the irreversible contributions are significant. $J_{\text{c}}(B)$ is taken from
Eq.~(\ref{eqn:JcBSim}) with $J_{\text{c0}} = 2 \times 10^8$\,Am$^{-2}$, $B_{\text{0}} = 0.02$\,T,
and $\alpha = 0.5$. The dashed line illustrates the pure reversible fraction - $M_{\rm
r}(H_{\text{a}})$ (obtained from the approximate equations of Ref.~\onlinecite{Bra03a}), where
$B_\text{c2} = 1$\,T and $\kappa = 4.8$ - thus $B_\text{c1} \simeq 0.05$\,T.}
\end{figure}

Step (ix) finishes the procedure, in case the reversible magnetization is not ignored, and we
can proceed with step (ii). Figure~ \ref{fig:mrevirr} presents an example of a magnetization loop,
where both the reversible and irreversible contributions are significant. For describing flux line
motion we employ
\begin{equation}
 	\label{eqn:jHElectric}
	\vec{E} = A_{\text{c}}\,B\,\dfrac{( J_\text{H} / J_{\text{c}} )^n}{1 + ( J_\text{H} /
J_{\text{c}}
)^n} \vec{J}_\text{H}
\end{equation}
as suggested in Ref.~\onlinecite{Bra99a}, which correctly leads to flux flow ($E \propto B\,
J_\text{H}$) at $J_\text{H} \gg J_\text{c}$ and flux creep at $J_\text{H}  \ll J_\text{c}$. Here,
$A_{\text{c}}$ is an appropriate constant, equivalent to $E_{\text{c}}$ of Eq.~(\ref{eqn:PowerLaw}),
but given in units of VmT$^{-1}$A$^{-1}$.

In the following, we mainly concentrate on samples with high critical current density, where the
reversible properties may be ignored and therefore only steps (i - vii) will be executed.

\subsection{Technical aspects}

\begin{figure}
    \centering
	 \includegraphics[clip, width = 5cm]{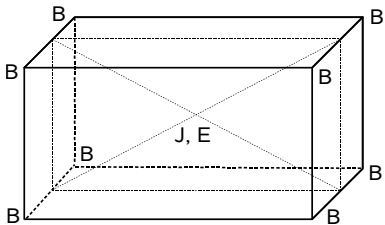}
    \caption{\label{fig:element} (Color online)  Schematic view of a single element, (in which
$\vec{J}$ is constant) and the positions, where $\vec{B}$, $\vec{E}$, and $\vec{J}$ are calculated.
The dotted square indicates the two dimensional case (see the text for details).}
\end{figure}

In this section, we address some specific points of our implementation. In most cases, we deal with
rectangular samples and, therefore, subdivide the sample into small rectangular elements, within
which the current density is kept constant. For simplicity, we work with an equidistant grid, but
note that a better efficiency is expected from a grid with smaller elements near the surface, since
most quantities typically exhibit larger variations there. Each grid element consists of eight
vertices and a central point (Fig.~ \ref{fig:element}). The sequence (Fig.~\ref{fig:flow}) dictates
the position, on which each quantity is evaluated, in a natural way  (Fig.~ \ref{fig:element}). The
induction $\vec{B}$ is defined at the vertices, which also include the sample surface. Accordingly,
its derivative (or actually its curl) is calculated at the midpoint position using $\vec{B}$ values
of the next eight adjacent vertices. The electric field is in most cases a direct function of the
currents (e.g. Eq.~\ref{eqn:PowerLaw}) and is therefore also calculated at the center of the
elements. In the next step, we need the curl of $\vec{E}$, which naturally leads us back to the
vertices and the results ($\partial_{\rm t} \vec{B}$) immediately allow modifying $\vec{B}$ at those
positions. Note that calculating the curl of $\vec{E}$ at the sample surface is not required.

At the sample surface, the induction is obtained by integration, which is approximated by a
summation over the discrete elements $k$ within which $J$ is  constant. The three components ($x$,
$y$, and $z$) of Eq.~(\ref{eqn:Hsurface}) may be written as
\begin{eqnarray}
 	\label{eqn:Hs_x}
	H_{\rm s, x}(\vec{r}_{\rm s}) &=& \sum_k \left[ J_{\rm y}^{\rm (k)} A_{\rm z}^{\rm
(k)}(\vec{r}_{\rm s})
- J_{\rm z}^{\rm (k)} A_{\rm y}^{\rm (k)}(\vec{r}_{\rm s})   \right]  \\
 	\label{eqn:Hs_y}
	H_{\rm s, y}(\vec{r}_{\rm s}) &=& \sum_k \left[ J_{\rm z}^{\rm (k)} A_{\rm x}^{\rm
(k)}(\vec{r}_{\rm s})
- J_{\rm x}^{\rm (k)}
A_{\rm z}^{\rm (k)}(\vec{r}_{\rm s})   \right]  \\
 	\label{eqn:Hs_z}
	H_{\rm s, z}(\vec{r}_{\rm s}) &=& \sum_k \left[ J_{\rm x}^{\rm (k)} A_{\rm y}^{\rm
(k)}(\vec{r}_{\rm s})
- J_{\rm y}^{\rm (k)}
A_{\rm x}^{\rm (k)}(\vec{r}_{\rm s})   \right] 
\end{eqnarray}
The sums run over all elements of the grid,  $J_{\rm i}^{\rm (k)}$ denotes the $i$-component
($i = x, y$, or $z$) of the current density of the $k^{\rm th}$ element, and $A_{\rm i}^{\rm
(k)}(\vec{r}_{\rm s})$ the $i$-component of the function
\begin{equation}
	\label{eqn:Aks}
	\vec{A}^{\rm (k)}(\vec{r}_{\rm s}) = \frac{1}{4 \pi} \int_{\rm (k)}\!d^3 {\rm r'}
\frac{\vec{r}_{\rm s}
- \vec{r}'}{|\vec{r}_{\rm s} - \vec{r}'|^3}
\end{equation}
where $\vec{r}_{\rm s}$ denotes a vertex position at the sample surface and the integral is
performed over the volume of the $k^{\rm th}$ element. We point out that it is essential to
calculate these integrals numerically very accurately. Note, however, that $\vec{A}^{\rm
(k)}(\vec{r}_{\rm s})$ depends only on the position vectors, hence the integrals need to be
calculated  just once during the initial stage, and can be stored for later use in
Eqs.~(\ref{eqn:Hs_x}-\ref{eqn:Hs_z}).

Finally, we address the problem of choosing an adequate time increment $\Delta t$, which should be
as large as possible to make the simulation time short, but small enough to have no influence on the
results. In our implementation, it turns out that choosing $\Delta t$ too large makes the
calculations unstable with $|\vec{J}|$ increasing rapidly towards infinity. This situation can be
easily monitored and mitigated by resuming the simulation from a previously saved configuration with
a smaller $\Delta t$. It turns out that even smaller values of $\Delta t$ do not modify the results.

\subsection{Magnetization loops of rectangular samples}

The model was successfully applied to fully 3D calculations in conveniently short simulation times,
as reported in Ref.~\onlinecite{Zeh06a}. In the following, we focus on magnetization loops of
rectangular samples in the presence of an homogeneous external field, which occurs in most
magnetometry experiments (e.g. SQUID or VSM). The external field is applied in $z$ direction,
perpendicular to the top and bottom surface of our sample, and we assume isotropic material
parameters (e.g. $J_{\text{c}}$) within the $xy$ plane. In this case, the model may be simplified
assuming the currents to flow exactly parallel to the nearest sample surfaces. The benefits of the
procedure are still available in the 2 dimensional calculations, making it possible to simulate
realistic magnetization loops very rapidly (e.g. within about 1 to 10 minutes on a conventional
standard PC). We point out that our assumption on the current flow is not rigorously valid (see
Ref.~\onlinecite{Bra95b}), but the deviations are only significant in very thin samples, which are
not addressed here.

For the actual implementation, we assume the sample center to coincide with the point of origin and
the lateral surfaces to be parallel to the $x$ or $y$ axis. The fields and currents  need to be
calculated only at the cross section spanned by the $z$ and $y$ axis, on which a 2D grid is defined.
Thus we apply almost the same procedure as illustrated in Fig. \ref{fig:element}, but mapped on 2
dimensional elements (i.e. the cross section of each 3D element - see Fig.~ \ref{fig:element}). The
calculation of the surface field (Eqs.~\ref{eqn:Hs_x} - \ref{eqn:Hs_z}) is slightly modified to
\begin{eqnarray}
 	\label{eqn:Hs2d_x}
	H_{\rm s, x}(\vec{r}_{\rm s}) &=& 0  \\
 	\label{eqn:Hs2d_y}
	H_{\rm s, y}(\vec{r}_{\rm s}) &=& \sum_m  J^{\rm (m)} C_{\rm y}^{\rm (m)}(\vec{r}_{\rm
s})  \\
 	\label{eqn:Hs2d_z}
	H_{\rm s, z}(\vec{r}_{\rm s}) &=& \sum_m J^{\rm (m)} C_{\rm z}^{\rm (m)}(\vec{r}_{\rm s})
\end{eqnarray}
with
\begin{equation}
	\label{eqn:Cms}
	\vec{C}^{\rm (m)}(\vec{r}_{\rm s}) = \frac{1}{4 \pi} \int_{\rm (m)}\!d^3 {\rm r'}
\frac{\vec{\epsilon}_{\rm J} \times ( \vec{r}_{\rm s} - \vec{r}')}{|\vec{r}_{\rm s} - \vec{r}'|^3}
\end{equation}
$H_{\rm s, x}(\vec{r}_{\rm s})$ vanishes due to the symmetry of the system. The sums run over all 2
dimensional elements ($m$) of the sample cross section. The integration is performed over one
element ($m$) and - for the third dimension - over the closed current loop, exactly parallel to the
nearest sample surfaces. $\vec{\epsilon}_{\rm J} $ indicates the unit vector of the a-priori known
current flow direction. Thus $\vec{C}^{\rm (m)}(\vec{r}_{\rm s})$ can also be accurately calculated
during the initial stage and stored for later use.

\subsection{Tests of the simulations}

\begin{figure}
    \centering
    \includegraphics[clip, width = 8.5cm]{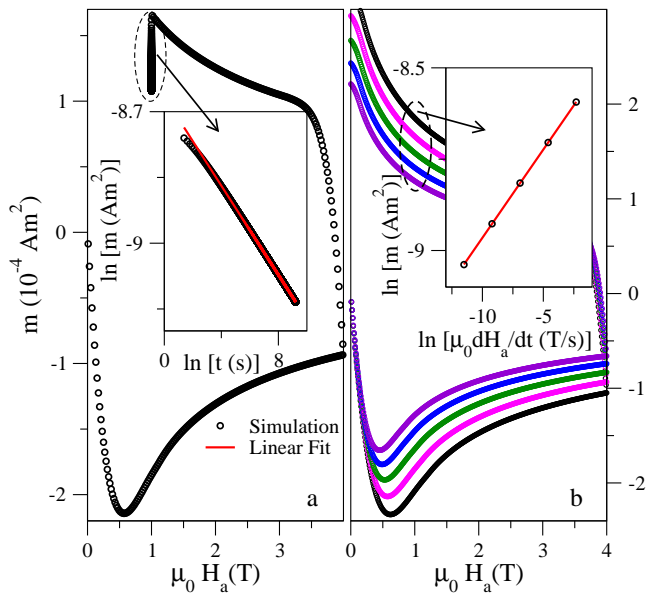}
    \caption{\label{fig:relax} (Color online)  Verification of the creep behavior in the
simulations. Panel (a) present a magnetization loop ($n$ = 20) followed by recording
relaxation at 1\,T for about 3 hours. The inset shows the relaxation rate ${\rm d} \ln m / {\rm d}
\ln t~ (\simeq -1 / n)$ together with a linear fit revealing $n = 19.7$. Panel (b) shows
magnetization loops at different field sweep rates (10$^{-5}$,10$^{-4}$,10$^{-3}$,10$^{-2}$,
and 10$^{-1}$ T/s) and the inset the dynamical relaxation rate ${\rm d} \ln m / {\rm d} \ln \mu_0
\dot{H}_{\rm a}~ (\simeq 1 / n)$ at 1\,T, where the linear fit leads to $n = 20.7$.}
\end{figure}

Several tests are available for verifying the output of the simulations. The first concerns
$H^\star$, the lowest applied field of a virgin $m(H_\text{a})$ curve, at which the currents have
penetrated the entire sample. It corresponds to the $z$-component of the field of that current
distribution at the center of the sample, which is  easily calculated for constant currents
(Bean model, e.g. Eq.~\ref{eqn:Hsurface}, but with $\vec{r} = (0, 0, 0)^{\rm T}$).

The simulations reveal $H^\star$ by monitoring the current density at the sample center.
The Bean model is approximated by applying constant currents and - to reduce relaxation - a large
$n$ value. As expected, the deviation of $H^\star$ from the direct calculation is reduced
with increasing $n$ value and almost vanishes (smaller than 1\%) at $n > 100$.

Secondly, we can test the creep behavior of the currents by studying the relaxation of the magnetic
moment. This time, the power law (Eq.~\ref{eqn:PowerLaw}) was applied with a field dependent
critical current density and a constant $n$ value ($n = 20$). Imitating the common experimental
procedure, we started with simulating (parts of) a magnetization loop, followed by recording the
relaxation in the fully penetrated state, i.e. the time dependence of the magnetic moment - $m(t)$ -
at constant applied field (Fig.~\ref{fig:relax}a). According to theory the $n$ value may be
extracted via $|{\rm d} \ln m / {\rm d} \ln t| \simeq 1 / n$ which matches very well ($\sim 1.5$\%
deviation) the input value of $n$ in the simulation (see inset of Fig.~\ref{fig:relax}a).

A second way of studying relaxation is provided by the so-called dynamic relaxation, where
magnetization loops are measured (or simulated) at different field sweep rates $\dot{H}_{\rm a} =
{\rm d} H_\text{a} / {\rm d} t$ (Fig.~\ref{fig:relax}b). The relaxation rate at constant applied
field is now given by $|{\rm d} \ln m / {\rm d} \ln \mu_0 \dot{H}_{\rm a}| \simeq 1 / n$ and is
again found to be in good agreement ($\sim 3.5$\% deviation) with our input value of $n$ (see inset
of Fig.~\ref{fig:relax}b).

\section{Application to magnetometry \label{sec:application}}

\subsection{Evaluation of the critical current density}

\subsubsection{Experiment \label{sec:experimentalEval}}

In the following we present two applications of the simulations to common problems of magnetometry
experiments. The first refers to the evaluation of the critical current density from magnetization
loops. Usually the Bean model is applied, which presumably leads to certain deviations from the
correct $J_{\text{c}}$ behavior. To analyze these deviations quantitatively we start from a given
$J_{\text{c}}(B)$ behavior and simulate magnetization loops. Then we apply the same evaluation
procedure as used for the experimental data to reevaluate $J_{\text{c}}$, which is finally compared
with our input $J_{\text{c}}$ curve.

We start by explaining the experimental procedure used in most experiments for acquiring
$J_{\text{c}}(H_\text{a})$ from magnetometry measurements and introduce a simple extension that
allows to obtain $J_{\text{c}}(B )$. When determining the critical current density from SQUID or VSM
measurements we usually do not know more than the magnetic moment as a function of applied field,
$m(H_\text{a})$, and the sample dimensions $a$, $b$, $c$, where we assume $c$ to be parallel to
$H_\text{a}$. Thus, assumptions on $J_{\text{c}}$ are inevitable, which are in most cases a constant
absolute value of $\vec{J}$ equal to $J_{\text{c}}$ and a flow direction parallel to the nearest
sample surfaces (Bean model). The relation between $m$ and $J_{\text{c}}$ is given by
Eq.~(\ref{eqn:moment}) leading to an analytic expression for rectangular samples in the fully
penetrated state\cite{Cam72a} that can be solved for $J_{\text{c}}$
\begin{equation}
	\label{eqn:JcBean}
	J_{\text{c}} = \frac{|m_{\text{i}}|}{\Omega} \frac{4}{b (1 - \frac{b}{3a})}~~~~~~~~~~{\rm with}~a
\geq b
\end{equation}
$\Omega = a b c$ and $m_{\text{i}}$ denotes the irreversible magnetic moment generated by the
critical currents, which is  given by half of the hysteresis width of the magnetization loop.
Equation~(\ref{eqn:JcBean}) leads to a first approximation of $J_{\text{c}}$ as a function of the
applied field - $J_{\text{c}}(H_\text{a})$. To come closer to the actual material property, i.e.,
the critical current density as a function of the magnetic induction - $J_{\text{c}}(B)$ - (which is
independent of sample geometry), we additionally consider the field induced by the current
distribution (cf. Ref.~\onlinecite{Wie92a} for the case of cylindrical samples). Since the magnetic
moment acquired by the magnetometry measurements always refers to the whole sample volume, the
induction needs to be averaged in the same way to obtain $m(B)$, i.e.
\begin{equation}
	\label{eqn:BGesamt}
	B = \mu_0 \langle |H_\text{a} +  H_{\rm s, z}| \rangle
	\end{equation} 
with $\vec{H}_{\text{s}}$ from Eq.~(\ref{eqn:Hsurface}), and
\begin{equation}
	\label{eqn:Hs_weighting}
	\langle |H_\text{a} +  H_{\rm s, z}| \rangle =  \frac{1}{w} \int \!d^3{\rm r}\, (\vec{r} \times
\vec{\epsilon}_{\rm J})_{\rm z}\, |H_\text{a} +  H_{\rm s, z}(\vec{r})|
\end{equation}
The integral runs over the whole sample volume. $(\vec{r} \times \vec{\epsilon}_{\rm J})_{\rm z}$
represents a weighting factor, which corresponds to that of $J_{\text{c}}$ in the calculation of $m$
(cf. Eq.~\ref{eqn:moment}) and reflects also the distance from the current element to the pick-up
coils of the magnetometry devices.  $w$ is the integral of this factor over the sample volume. The
weighting does not strongly affect the results, but is still significant for the behavior of
$J_{\text{c}}(B)$. Note further, that we evaluate $J_{\text{c}}$ as a function of the absolute value
of the $B$ component parallel to the applied field ($B_{\rm z}$).

Finally, we address the evaluation of $m_{\text{i}}$. The experiments provide $m(H_\text{a})$, which
is usually the sum of $m_{\text{i}}$ and of the reversible magnetic moment $m_{\text{r}}$.
Additional magnetic signals from the sample or the sample holder can usually be treated in a similar
way as $m_{\text{r}}$. $m_{\rm i}$ is extracted from (half of) the difference of $m$ at increasing
($H_{\rm +}$, $B_{\rm +}$) and decreasing ($H_{\rm -}$, $B_{\rm -}$) fields at the same induction
$B$. But note that the two branches of $m(H_\text{a})$ refer to different values of $B$ at the same
$H_\text{a}$ since the currents have opposite orientation in the two branches. Two situations are
distinguished. First, when the irreversible part dominates, as is the case in most experiments, $B$
is separately calculated for each branch of $m(H_\text{a})$ by applying Eqs.~(\ref{eqn:JcBean} -
\ref{eqn:BGesamt}) with $m_{\rm i} \simeq m$. Then $m_{\text{i}}(B)$ is obtained from
\begin{equation}
	\label{eqn:mirr}
	m_{\text{i}}(B) = \frac{m(B_-) - m(B_+)}{2}
\end{equation} 
and the final $J_{\text{c}}(B)$ by evaluating Eq.~(\ref{eqn:JcBean}) again. In the second case,
where the reversible parts are significant or even dominate, the critical current density is usually
low and therefore the corresponding field correction ($H_{\rm s, z}$) small. Accordingly, we start
with evaluating $m_{\text{i}}(H_\text{a}) = 0.5 [m(H_{\rm a -}) - m(H_{\rm a +})]$, i.e. the
irreversible magnetic moment from the hysteresis width at the same applied field, and then apply
Eqs.~(\ref{eqn:JcBean} - \ref{eqn:BGesamt}) for getting $J_{\text{c}}(B)$.

\subsubsection{Simulation}

\begin{figure}
    \centering
	\includegraphics[clip, width = 8.5cm]{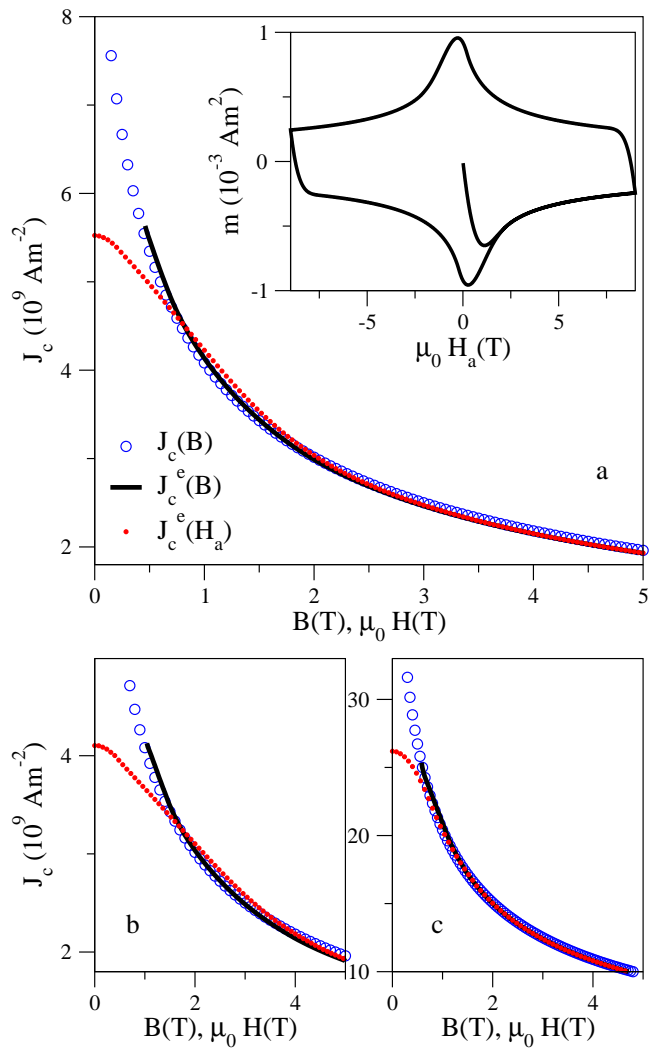}
    \caption{\label{fig:Jcs} (Color online)  Results for different sample geometries: $a \times b
\times c = 1 \times 1 \times 1$\,mm$^3$ in (a), $3 \times 3 \times 3$\,mm$^3$ in (b), and $1 \times
0.5 \times 0.1$\,mm$^3$ in (c). The open symbols show the input $J_{\text{c}}(B)$ used to simulate
the magnetization loops (e.g. inset of a), the dotted curve $J_{\text{c}}(H_\text{a})$ and the solid
line $J_{\text{c}}(B)$ evaluated from the magnetization curves by using Bean's model (see text).}
\end{figure}

The above experimental evaluation is not expected to result in the ``true'' $J_{\text{c}}(B)$ curve.
The main reason for discrepancies is the (necessary) assumption of a constant current density within
the sample, which is not fulfilled in realistic materials, since $B$ may vary considerably and
$J_{\text{c}}(B)$ is usually not constant. Accordingly, we expect more pronounced deviations, when
the samples are large and have high $J_{\text{c}}$, since both properties would enhance the (peak to
peak) variation of $B$ within the sample. At low fields this variation is comparable with $\mu_0
H^\star$, which may amount to several Tesla in typical ``SQUID or VSM'' samples. At higher fields,
we expect better agreement with the ``true'' $J_{\text{c}}$ , since $J_{\text{c}}$ becomes smaller
and flatter as a function of $B$ in most cases.

To get some idea of the quantitative differences between the ``true''  $J_{\text{c}}$ and
$J_{\text{c}}$ from the above evaluation, we simulated a typical magnetometry measurement assuming a
sample size of $1 \times 1 \times 1$\,mm$^{\rm 3}$, and a $J_{\text{c}}$ behavior described by
\begin{equation}
	\label{eqn:JcBSim}
	J_{\text{c}}(B) = \frac{J_{\rm c0}}{(1 + B / B_0)^\alpha}
\end{equation}
Setting $B_0 = 0.2$\,T and $\alpha = 1/2$ roughly leads to a behavior observed in many (melt
textured) Y-123 (Y$_1$Ba$_2$Cu$_3$O$_{7-x}$) samples at low temperatures. $J_{\rm c0}$ = $1 \times
10^{10}$\,Am$^{-2}$ leads to $\mu_0 H^\star \simeq 2.1$\,T and a similarly large variation of $B$ in
the sample.

Furthermore, we applied the power law (Eq.~\ref{eqn:PowerLaw}) with $E_{\rm c} = 5 \times
10^{-6}$\,Vm$^{-1}$ and $n = 20$ and a field sweep rate of $\mu_0 \dot{H}_{\rm a} = 1 \times
10^{-2}$\,Ts$^{-1}$. This rather fast field sweep rate corresponds to the maximum we can set in our
VSM measurements and was chosen to reduce relaxation effects, i.e. to get $J \simeq J_{\text{c}}$,
where $J$ is the real flowing current.

The inset of Fig.~\ref{fig:Jcs}a presents the simulated magnetization curve of this sample up to
fields of 9\,T, including the virgin curve. Before applying the methods of section
\ref{sec:experimentalEval}, all parts of the curve, where the currents do not flow in the whole
sample or do not have the same orientation need to be removed. This is done automatically by the
evaluation program and includes the appropriate parts of the virgin curve (i.e. for $H_\text{a} <
H^\star$) and those parts directly after reversing the field, where the currents have opposite
orientation.

The results are illustrated in Fig.~\ref{fig:Jcs}a. The open symbols show the original (input)
$J_{\text{c}}(B)$, i.e. the ``true'' $J_{\text{c}}(B)$,  according to Eq.~(\ref{eqn:JcBSim}). The
dotted line presents $J_{\text{c}}^{\text{e}}(H_\text{a})$, where the subscript "e" indicates that
$J_{\text{c}}$ has been evaluated from the magnetization curve (of the inset of Fig.~\ref{fig:Jcs}a)
with the (experimental) methods of section \ref{sec:experimentalEval}, and the solid line is the
final result $J_{\text{c}}^{\text{e}}(B)$  when considering also the field correction according to
Eq.~(\ref{eqn:BGesamt}). $J_{\text{c}}^{\text{e}}(H_a)$ shows significant deviations from
$J_{\text{c}}(B)$ at fields $H_\text{a} < H^\star$. $J_{\text{c}}^{\text{e}}(B)$, however, matches
quite well despite the rather rough assumption of a constant current density in the sample (note
that the peak to peak variation of $B$ in the sample reaches 2.4 \,T at constant applied field in
this specific simulation). Slower field sweep rates or field-step measurements would shift the
curves to slightly smaller values due to relaxation. Note further that the evaluation method does
not give access to $J_{\text{c}}$ at or close to $B = 0$, since the mean induction (cf.
Eq.~\ref{eqn:Hs_weighting}) is always larger than zero.

To get further insight, we repeated the procedure for different samples and $J_{\text{c}}(B)$ curves
(e.g. with exponential behavior) and found similarly good agreement as above. For instance,
Fig.~\ref{fig:Jcs}b shows results on a sample with dimensions $3 \times 3 \times 3$\,mm$^{\rm 3}$
and the same $J_{\text{c}}(B)$ as in the above example (only $E_{\rm c}$ is slightly adapted to
account for the larger sample size). Although  $\mu_0 H^\star$ of 4.4\,T is quite large and
$J_{\text{c}}^{\text{e}}(H_a)$ starts to deviate from $J_{\text{c}}(B)$ at a similarly high field,
these deviations are again nicely corrected by $J_{\text{c}}^{\text{e}}(B)$.

Analyzing situations with much larger $\mu_0 H^\star$ does not make sense, since the evaluation of
$J_{\text{c}}$ requires external fields larger than $\mu_0 H^\star$, but typical experimental
devices do not provide fields much above 5 - 10\,T. Increasing $J_{\text{c}}(B)$ instead of the
sample size leads to similar effects.

As a final example, we show results on a sample with a size of  $1 \times 0.5 \times 0.1$\,mm$^{\rm
3}$, which is representative for a typical single crystal geometry. We took the same
$J_{\text{c}}(B)$ behavior as in the above examples but with $J_{\rm c0}$ = $5 \times
10^{10}$\,Am$^{-2}$, accordingly $\mu_0 H^\star \simeq 2.1$\,T. Fig.~\ref{fig:Jcs}c presents the
results, which again demonstrate the excellent agreement between the evaluated
$J_{\text{c}}^{\text{e}}(B)$ and the input curve $J_{\text{c}}(B)$.

\subsection{Superconducting volume fraction}

The evaluation of the critical current density as described in the previous sections is only valid
if the current flow is unimpeded over the whole sample. This is illustrated by
Eq.~(\ref{eqn:JcBean}), which shows that a sample with dimensions $a \times b \times c$ would induce
a larger magnetic moment than the sum of two samples with dimensions $a \times b/2 \times c$, with
the same $J_{\text{c}}$ as the large sample. Thus, grain boundaries and other macroscopic
inhomogeneities (e.g. normal conducting inclusions), which impede the current flow,
would lead to an underestimation of $J_{\text{c}}$, when applying the above method. A similar effect
is caused by overestimating the superconducting sample volume, e.g. when the material properties are
degraded at the surface. We further note, that magnetization measurements are not only used for
evaluating $J_{\text{c}}$, but also for determining all kinds of superconducting parameters, such as
the critical magnetic fields or the characteristic lengths, etc. (e.g. Ref.~ \onlinecite{Zeh02a}).
Most of these quantities are derived from the magnetization, $M = m / \Omega$, i.e., it is again
important to exactly know the volume ($\Omega$). We point out that both surface degradation and
grain boundaries were even found in samples believed to be single crystals. In the following we show
that the reverse branch of a magnetization loop is very sensitive to such imperfections, and that we
can prove whether or not a sample is single grained, and even detect, if parts of the surface are
degraded.

\begin{figure}
    \centering
	\includegraphics[clip, width = 8.5cm]{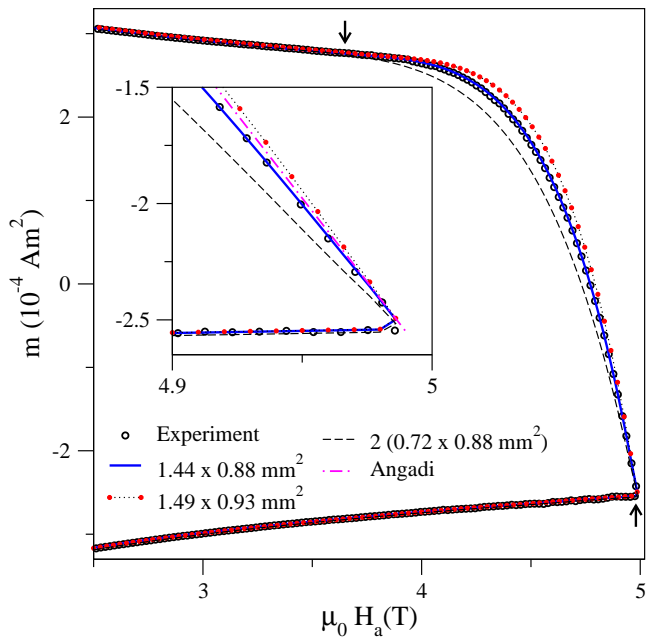}
    \caption{\label{fig:revleg} (Color online)  Comparison of the magnetic moment from experiment
and simulations with different sample surface sizes. The arrows indicate the interval, within which
currents with opposite orientation flow in the sample.}
\end{figure}

We started our analysis with experimental magnetization loops of a Y-123 sample at 30\,K. The size
of the rectangular sample was precisely measured with an optical microscope and found to be $a
\simeq 1.44$\,mm, $b \simeq 0.88$\,mm, and $c \simeq 0.76$\,mm. Using these data, $J_{\text{c}}(B)$
was calculated with the above method (section \ref{sec:experimentalEval}), and the result was used
for simulating the magnetization loops. As expected, there is agreement with experiment over most
part of the field range. Small differences are only observed near $H_\text{a} = 0$, since
$J_{\text{c}}$ close to $B = 0$ needs to be extrapolated. We are, however, not interested in this
region, but look closer to that part of the reverse branch of the loop, where currents with opposite
orientation flow in the sample (indicated by the two  arrows in Fig. \ref{fig:revleg}) and which had
been disregarded for evaluating $J_{\text{c}}$. As shown in Fig.~\ref{fig:revleg} there is perfect
agreement between experiment (open circles) and simulation (solid line, $a \times b = 1.44 \times
0.88$\,mm) even in this part of the loop, thus demonstrating that the size of the sample surface had
been determined correctly and that the sample is pretty homogeneous for the macroscopic current
flow. Comparing only the magnetization slope directly after the reversal point (up-arrow in
Fig.~\ref{fig:revleg}) (inset of Fig.~\ref{fig:revleg}) provides information on the sample size and
surface, but not on the interior since the newly penetrating currents (which have opposite
orientation with respect to the existing currents) flow only at the sample surface. With decreasing
applied field, the new currents penetrate into the sample and as soon as the current front reaches a
region with a different $J_{\text{c}}$ (e.g. at a grain boundary or a normal conducting inclusion)
the macroscopic current path and therefore also the magnetic moment would start to deviate from the
simulations (where perfect homogeneity is assumed). In addition, the field, where the new currents
arrive at the center (down-arrow in Fig.~\ref{fig:revleg}), would be different. To conclude, only
comparing experiment and simulation over the whole penetration interval (i.e. form the up- to the
down-arrow in Fig.~\ref{fig:revleg}) can clarify if samples are homogeneous (at least for the
current flow) over the whole volume.

Let us assume that the measurement of the sample size had resulted in slightly larger values, namely
$a = 1.49$\,mm and $b = 0.93$\,mm, i.e. only 0.05\,mm for both lengths, which increases the volume
by about 9\%. We repeated the whole procedure with these data (which leads to a slightly smaller
$J_{\text{c}}$) and find again agreement on most parts of the $m(H_\text{a})$ loops. However,
inspecting the reverse branch more closely, we find significant disagreement between the two arrows
(Fig. \ref{fig:revleg}), i.e. where not all currents are equally oriented, which demonstrates
different sample sizes. This indicates either an error in the measurement of the sample size, or the
fact that small regions of the lateral surfaces are non-superconducting. While a smaller
$J_{\text{c}}$ hardly affects the slope of $m(H_\text{a})$ close to the reversal point, the larger
sample size and the correspondingly larger current loop enhances the (absolute value of the) slope
by about 10\% as shown in the inset of Fig. \ref{fig:revleg} (note that $H^\star$ also increases).

Secondly, we studied the case of a sample divided into two independent grains (bicrystal) with sizes
$0.72 \times 0.88 \times 0.76$\,mm$^3$ (i.e. $a/2 \times b \times c$). Since the
relative contribution of the critical currents to the magnetic field is small at high applied
field,  any interaction between the two grains (i.e. magnetic fields from the other grain) was
neglected. Having the same overall magnetic moment (which is simply the sum of $m$ from the two
grains) leads to a larger $J_{\text{c}}$ than in the large (single grained) sample. As expected, the
slope of $m(H_\text{a}$) at the reversal point is significantly flatter in the bicrystal (about
17\%) than in the single grain sample (inset of Fig. \ref{fig:revleg}) making it easy to distinguish
between the different situations.

Finally, we compare our results on the slope of the reverse branch with the well known analytical
expression given by Angadi et al\cite{Ang91a}, i.e. d$\text{m} / \text{d}H = -\pi^2 R^3 / \Theta$,
with $\Theta = \ln{(8 R / c)} - 0.5$ and $R$ the radius of a cylinder which has the same surface
area as the cuboid. Using this method results in a slope that deviates from simulation and
experiment by only about 5\% (inset of Fig. \ref{fig:revleg}). However, applying the same method
to the bicrystal leads to a slope that actually matches the exact result of the single grain sample
even better. Thus, this approximation cannot be recommended to verify if samples are single
grained or decomposed into a few grains.

\section{Summary}

Details of a simple method for two or three dimensional simulations of the current dynamics in a
superconductor have been presented. Solving the differential form of Maxwell's equations in the
sample interior makes the calculations very fast. The integral form is only used at the sample
surfaces to meet the correct boundary conditions. The procedure results in the time and position
dependence of the magnetic induction and the critical current density, which makes it very
convenient to apply field dependent material parameters, like $J_{\text{c}}(B)$. The method allows
simulating different kinds of experiments, like magnetization loops, static and dynamic relaxation
measurements, etc.,  in which also the reversible superconducting properties may be considered.

To demonstrate the potential of the method, we presented two examples, which are important for
standard magnetometry measurements. In the first, magnetization loops were simulated by applying a
given $J_{\text{c}}(B)$ and then evaluated by the methods usually applied to experimental data, i.e.
the Bean model. Comparison of the given $J_{\text{c}}(B)$ and the 'Bean' $J_{\text{c}}(B)$ shows
good agreement even for large samples and high $J_{\text{c}}$, if the 'Bean' result is corrected by
a simple numerical expression.

Secondly, we showed that surface degradations, which reduce the superconducting volume fraction of a
sample, or inhomogeneities like grain boundaries, which influence the current flow, can be detected
by carefully comparing the magnetization loops from experiment and simulation at regions, where
currents with opposite orientation flow in the sample.

\begin{acknowledgments}
I wish to thank Franz Sauerzopf and Harald W. Weber for helpful discussions.
This work was supported by the Austrian Science Fund under contract 21194.
\end{acknowledgments}


\begin{thebibliography}{18}
\expandafter\ifx\csname natexlab\endcsname\relax\def\natexlab#1{#1}\fi
\expandafter\ifx\csname bibnamefont\endcsname\relax
  \def\bibnamefont#1{#1}\fi
\expandafter\ifx\csname bibfnamefont\endcsname\relax
  \def\bibfnamefont#1{#1}\fi
\expandafter\ifx\csname citenamefont\endcsname\relax
  \def\citenamefont#1{#1}\fi
\expandafter\ifx\csname url\endcsname\relax
  \def\url#1{\texttt{#1}}\fi
\expandafter\ifx\csname urlprefix\endcsname\relax\def\urlprefix{URL }\fi
\providecommand{\bibinfo}[2]{#2}
\providecommand{\eprint}[2][]{\url{#2}}

\bibitem[{\citenamefont{Prigozhin}(1996)}]{Pri96a}
\bibinfo{author}{\bibfnamefont{L.}~\bibnamefont{Prigozhin}},
  \bibinfo{journal}{J. Comput. Phys.} \textbf{\bibinfo{volume}{129}},
  \bibinfo{pages}{190 } (\bibinfo{year}{1996}).

\bibitem[{\citenamefont{Brandt}(1996)}]{Bra96a}
\bibinfo{author}{\bibfnamefont{E.~H.} \bibnamefont{Brandt}},
  \bibinfo{journal}{Phys. Rev. B} \textbf{\bibinfo{volume}{54}},
  \bibinfo{pages}{4246} (\bibinfo{year}{1996}).

\bibitem[{\citenamefont{Brandt}(1999)}]{Bra99a}
\bibinfo{author}{\bibfnamefont{E.~H.} \bibnamefont{Brandt}},
  \bibinfo{journal}{Phys. Rev. B} \textbf{\bibinfo{volume}{59}},
  \bibinfo{pages}{3369} (\bibinfo{year}{1999}).

\bibitem[{\citenamefont{Labusch and Doyle}(1997)}]{Lab97a}
\bibinfo{author}{\bibfnamefont{R.}~\bibnamefont{Labusch}} \bibnamefont{and}
  \bibinfo{author}{\bibfnamefont{T.~B.} \bibnamefont{Doyle}},
  \bibinfo{journal}{Physica C} \textbf{\bibinfo{volume}{290}},
  \bibinfo{pages}{143} (\bibinfo{year}{1997}).

\bibitem[{\citenamefont{Sanchez and Navau}(2001)}]{San01b}
\bibinfo{author}{\bibfnamefont{A.}~\bibnamefont{Sanchez}} \bibnamefont{and}
  \bibinfo{author}{\bibfnamefont{C.}~\bibnamefont{Navau}},
  \bibinfo{journal}{Phys. Rev. B} \textbf{\bibinfo{volume}{64}},
  \bibinfo{pages}{214506} (\bibinfo{year}{2001}).

\bibitem[{\citenamefont{Bad\'ia and L\'opez}(2002)}]{Bad02a}
\bibinfo{author}{\bibfnamefont{A.}~\bibnamefont{Bad\'ia}} \bibnamefont{and}
  \bibinfo{author}{\bibfnamefont{C.}~\bibnamefont{L\'opez}},
  \bibinfo{journal}{Phys. Rev. B} \textbf{\bibinfo{volume}{65}},
  \bibinfo{pages}{104514} (\bibinfo{year}{2002}).

\bibitem[{\citenamefont{Wolsky and Campbell}(2008)}]{Wol08}
\bibinfo{author}{\bibfnamefont{A.~M.} \bibnamefont{Wolsky}} \bibnamefont{and}
  \bibinfo{author}{\bibfnamefont{A.~M.} \bibnamefont{Campbell}},
  \bibinfo{journal}{Supercond. Sci. Technol.} \textbf{\bibinfo{volume}{21}},
  \bibinfo{pages}{075021} (\bibinfo{year}{2008}).

\bibitem[{\citenamefont{Campbell}(2009)}]{Cam09a}
\bibinfo{author}{\bibfnamefont{A.~M.} \bibnamefont{Campbell}},
  \bibinfo{journal}{Supercond. Sci. Technol.} \textbf{\bibinfo{volume}{22}},
  \bibinfo{pages}{034005} (\bibinfo{year}{2009}).

\bibitem[{\citenamefont{Lu et~al.}(2008)\citenamefont{Lu, Wang, Wang, and
  Zheng}}]{Lu08a}
\bibinfo{author}{\bibfnamefont{Y.}~\bibnamefont{Lu}},
  \bibinfo{author}{\bibfnamefont{J.}~\bibnamefont{Wang}},
  \bibinfo{author}{\bibfnamefont{S.}~\bibnamefont{Wang}}, \bibnamefont{and}
  \bibinfo{author}{\bibfnamefont{J.}~\bibnamefont{Zheng}}, \bibinfo{journal}{J
  Supercond Nov Magn} \textbf{\bibinfo{volume}{21}}, \bibinfo{pages}{467}
  (\bibinfo{year}{2008}).

\bibitem[{\citenamefont{Zehetmayer et~al.}(2006)\citenamefont{Zehetmayer,
  Eisterer, and Weber}}]{Zeh06a}
\bibinfo{author}{\bibfnamefont{M.}~\bibnamefont{Zehetmayer}},
  \bibinfo{author}{\bibfnamefont{M.}~\bibnamefont{Eisterer}}, \bibnamefont{and}
  \bibinfo{author}{\bibfnamefont{H.~W.} \bibnamefont{Weber}},
  \bibinfo{journal}{Supercond. Sci. Technol.} \textbf{\bibinfo{volume}{19}},
  \bibinfo{pages}{S429} (\bibinfo{year}{2006}).

\bibitem[{\citenamefont{Mikitik and Brandt}(2005)}]{Mik05a}
\bibinfo{author}{\bibfnamefont{G. P.}~\bibnamefont{Mikitik}} \bibnamefont{and}
  \bibinfo{author}{\bibfnamefont{E. H.}~\bibnamefont{Brandt}},
  \bibinfo{journal}{Phys. Rev. B} \textbf{\bibinfo{volume}{71}},
  \bibinfo{pages}{012510} (\bibinfo{year}{2005}).

\bibitem[{\citenamefont{Brandt and Mikitik}(2007)}]{Bra07a}
\bibinfo{author}{\bibfnamefont{E. H.}~\bibnamefont{Brandt}} \bibnamefont{and}
  \bibinfo{author}{\bibfnamefont{G. P.}~\bibnamefont{Mikitik}},
  \bibinfo{journal}{Phys. Rev. B} \textbf{\bibinfo{volume}{76}},
  \bibinfo{pages}{064526} (\bibinfo{year}{2007}).

\bibitem[{\citenamefont{Blatter et~al.}(1994)\citenamefont{Blatter, Feigel'man,
  Geshkenbein, Larkin, and M.Vinokur}}]{Bla94a}
\bibinfo{author}{\bibfnamefont{G.}~\bibnamefont{Blatter}},
  \bibinfo{author}{\bibfnamefont{M.~V.} \bibnamefont{Feigel'man}},
  \bibinfo{author}{\bibfnamefont{V.~B.} \bibnamefont{Geshkenbein}},
  \bibinfo{author}{\bibfnamefont{A.~I.} \bibnamefont{Larkin}},
  \bibnamefont{and}
  \bibinfo{author}{\bibfnamefont{V.}~\bibnamefont{M.Vinokur}},
  \bibinfo{journal}{Rev. Mod. Phys.} \textbf{\bibinfo{volume}{66}}
  (\bibinfo{year}{1994}).

\bibitem[{\citenamefont{Brandt}(1995{\natexlab{a}})}]{Bra95a}
\bibinfo{author}{\bibfnamefont{E.~H.} \bibnamefont{Brandt}},
  \bibinfo{journal}{Rep. Prog. Phys.} \textbf{\bibinfo{volume}{58}},
  \bibinfo{pages}{1465} (\bibinfo{year}{1995}{\natexlab{a}}).

\bibitem[{\citenamefont{Yeshurun et~al.}(1996)\citenamefont{Yeshurun,
  Malozemoff, and Shaulov}}]{Yes96a}
\bibinfo{author}{\bibfnamefont{Y.}~\bibnamefont{Yeshurun}},
  \bibinfo{author}{\bibfnamefont{A.~P.} \bibnamefont{Malozemoff}},
  \bibnamefont{and} \bibinfo{author}{\bibfnamefont{A.}~\bibnamefont{Shaulov}},
  \bibinfo{journal}{Rev. Mod. Phys.} \textbf{\bibinfo{volume}{68}},
  \bibinfo{pages}{911} (\bibinfo{year}{1996}).

\bibitem[{\citenamefont{Bean}(1962)}]{Bea62a}
\bibinfo{author}{\bibfnamefont{C.~P.} \bibnamefont{Bean}},
  \bibinfo{journal}{Phys. Rev. Lett.} \textbf{\bibinfo{volume}{8}},
  \bibinfo{pages}{250} (\bibinfo{year}{1962}).

\bibitem[{\citenamefont{Brandt}(2003)}]{Bra03a}
\bibinfo{author}{\bibfnamefont{E.~H.} \bibnamefont{Brandt}},
  \bibinfo{journal}{Phys. Rev. B} \textbf{\bibinfo{volume}{68}},
  \bibinfo{pages}{054506} (\bibinfo{year}{2003}).

\bibitem[{\citenamefont{Brandt}(1995{\natexlab{b}})}]{Bra95b}
\bibinfo{author}{\bibfnamefont{E.~H.} \bibnamefont{Brandt}},
  \bibinfo{journal}{Phys. Rev. B} \textbf{\bibinfo{volume}{52}},
  \bibinfo{pages}{15442} (\bibinfo{year}{1995}{\natexlab{b}}).

\bibitem[{\citenamefont{Campbell and Evetts}(1972)}]{Cam72a}
\bibinfo{author}{\bibfnamefont{A. M.}~\bibnamefont{Campbell}} \bibnamefont{and}
  \bibinfo{author}{\bibfnamefont{J. E.}~\bibnamefont{Evetts}},
  \bibinfo{journal}{Adv. Phys.} \textbf{\bibinfo{volume}{21}},
  \bibinfo{pages}{199} (\bibinfo{year}{1972}).

\bibitem[{\citenamefont{Wiesinger et~al.}(1992)\citenamefont{Wiesinger,
  Sauerzopf, and Weber}}]{Wie92a}
\bibinfo{author}{\bibfnamefont{H.}~\bibnamefont{Wiesinger}},
  \bibinfo{author}{\bibfnamefont{F.}~\bibnamefont{Sauerzopf}},
  \bibnamefont{and} \bibinfo{author}{\bibfnamefont{H.}~\bibnamefont{Weber}},
  \bibinfo{journal}{Physica C} \textbf{\bibinfo{volume}{203}},
  \bibinfo{pages}{121 } (\bibinfo{year}{1992}).

\bibitem[{\citenamefont{Zehetmayer et~al.}(2002)\citenamefont{Zehetmayer,
  Eisterer, Jun, Kazakov, Karpinski, Wisniewski, and Weber}}]{Zeh02a}
\bibinfo{author}{\bibfnamefont{M.}~\bibnamefont{Zehetmayer}},
  \bibinfo{author}{\bibfnamefont{M.}~\bibnamefont{Eisterer}},
  \bibinfo{author}{\bibfnamefont{J.}~\bibnamefont{Jun}},
  \bibinfo{author}{\bibfnamefont{S.~M.} \bibnamefont{Kazakov}},
  \bibinfo{author}{\bibfnamefont{J.}~\bibnamefont{Karpinski}},
  \bibinfo{author}{\bibfnamefont{A.}~\bibnamefont{Wisniewski}},
  \bibnamefont{and} \bibinfo{author}{\bibfnamefont{H.~W.} \bibnamefont{Weber}},
  \bibinfo{journal}{Phys. Rev. B} \textbf{\bibinfo{volume}{66}},
  \bibinfo{pages}{052505} (\bibinfo{year}{2002}).

\bibitem[{\citenamefont{Angadi et~al.}(1991)\citenamefont{Angadi, Caplin,
  Laverty, and Shen}}]{Ang91a}
\bibinfo{author}{\bibfnamefont{M.}~\bibnamefont{Angadi}},
  \bibinfo{author}{\bibfnamefont{A.}~\bibnamefont{Caplin}},
  \bibinfo{author}{\bibfnamefont{J.}~\bibnamefont{Laverty}}, \bibnamefont{and}
  \bibinfo{author}{\bibfnamefont{Z.}~\bibnamefont{Shen}},
  \bibinfo{journal}{Physica C} \textbf{\bibinfo{volume}{177}},
  \bibinfo{pages}{479 } (\bibinfo{year}{1991}).

\end{thebibliography}
\end{document}